\documentclass[sigconf,screen]{acmart}

\AtBeginDocument{%
  \providecommand\BibTeX{{%
    Bib\TeX}}}

\setcopyright{acmcopyright}
\copyrightyear{2024}
\acmYear{2024}
\acmDOI{XXXXXXX.XXXXXXX}

\acmConference[ICPC 2024]{32nd International Conference on Program Comprehension}{April 2024}{Lisbon, Portugal}
\acmPrice{15.00}
\acmISBN{978-1-4503-XXXX-X/18/06}

\usepackage{algorithmic}
\usepackage{graphicx}
\usepackage{tabularx}
\usepackage{textcomp}
\usepackage{adjustbox}
\usepackage{caption}
\usepackage{xcolor}
\usepackage{xspace}
\usepackage{fontawesome}
\usepackage{multirow}
\usepackage{rotating}
\usepackage{tikz}
\usepackage{soul}
\usepackage{makecell}
\usepackage{url}
\usepackage{enumitem}
\usepackage{booktabs}
\usepackage{fontawesome}

\def\BibTeX{{\rm B\kern-.05em{\sc i\kern-.025em b}\kern-.08em
    T\kern-.1667em\lower.7ex\hbox{E}\kern-.125emX}}

\newcommand{\ie}{\emph{i.e.,}\xspace}
\newcommand{\eg}{\emph{e.g.,}\xspace}

\newcommand{\etal}{\emph{et~al.}\xspace}
\newcommand{\secref}[1]{Section~\ref{#1}\xspace}

\newcommand{\figref}[1]{Fig.~\ref{#1}\xspace}

\newcommand{\tabref}[1]{Table~\ref{#1}\xspace}

\definecolor{Gray}{gray}{0.9}
\definecolor{codegreen}{rgb}{0,0.6,0}
\definecolor{codegray}{rgb}{0.73,0.38,0.06}
\definecolor{codepurple}{rgb}{0.70,0.27,0}
\definecolor{codemagenta}{rgb}{0.74,0.09,0.42}
\definecolor{codeoutput}{rgb}{0.5,0,0}
\definecolor{backcolour}{rgb}{0.96,0.96,0.96}

\newboolean{showcomments}

\setboolean{showcomments}{true}

\ifthenelse{\boolean{showcomments}}
{\newcommand{\nb}[2]{
		\fbox{\bfseries\sffamily\scriptsize#1}
		{\sf\small$\blacktriangleright$\textit{#2}$\blacktriangleleft$}
	}
	
}
{\newcommand{\nb}[2]{}
	
}

\def\BibTeX{{\rm B\kern-.05em{\sc i\kern-.025em b}\kern-.08em
    T\kern-.1667em\lower.7ex\hbox{E}\kern-.125emX}}

\begin{document}

\title[Towards Summarizing Code Snippets Using Pre-Trained Transformers]{Towards Summarizing Code Snippets\\Using Pre-Trained Transformers}

%\author{Anonymous Authors\\
%Anonymous Institution\\
%}

\author{Antonio Mastropaolo}
\affiliation{%
	\institution{SEART @ Software Institute, \\Universit\`a della Svizzera Italiana}
	\city{Lugano}
	\state{Switzerland}
	\country{CH}
}

\author{Matteo Ciniselli}
\affiliation{%
	\institution{SEART @ Software Institute, \\Universit\`a della Svizzera Italiana}
	\city{Lugano}
	\state{Switzerland}
	\country{CH}
}

\author{Luca Pascarella}
%\authornotemark[1]
\affiliation{%
	\institution{Center for Project-Based Learning, ETH Zurich}
	\city{Zurich}
	\state{Switzerland}
	\country{CH}
}

\author{Rosalia Tufano}
%\authornotemark[1]
\affiliation{%
	\institution{SEART @ Software Institute, \\Universit\`a della Svizzera Italiana}
	\city{Lugano}
	\state{Switzerland}
	\country{CH}
}

\author{Emad Aghajani}
%\authornotemark[1]
\affiliation{%
	\institution{SEART @ Software Institute, \\Universit\`a della Svizzera Italiana}
	\city{Lugano}
	\state{Switzerland}
	\country{CH}
}

\author{Gabriele Bavota}
%\authornotemark[1]
\affiliation{%
	\institution{SEART @ Software Institute, \\Universit\`a della Svizzera Italiana}
	\city{Lugano}
	\state{Switzerland}
	\country{CH}
}

\begin{abstract}
When comprehending code, a helping hand may come from the natural language comments documenting it that, unfortunately, are not always there. To support developers in such a scenario, several techniques have been presented to automatically generate natural language summaries for a given code. Most recent approaches exploit deep learning (DL)  to automatically document classes or functions, while little effort has been devoted to more fine-grained documentation (\eg documenting code snippets or even a single statement). Such a design choice is dictated by the availability of training data: For example, in the case of Java, it is easy to create datasets composed of pairs $<$$method$, $javadoc$$>$ that can be fed to DL models to teach them how to summarize a method. Such a comment-to-code linking is instead non-trivial when it comes to inner comments documenting a few statements. In this work, we take all the steps needed to train a DL model to automatically document code snippets. First, we manually built a dataset featuring 6.6k comments that have been (i) classified based on their type (\eg code summary, TODO), and (ii) linked to the code statements they document. Second, we used such a dataset to train a multi-task DL model taking as input a comment and being able to (i) classify whether it represents a ``code summary'' or not, and (ii) link it to the code statements it documents. Our model identifies code summaries with 84\% accuracy and is able to link them to the documented lines of code with recall and precision higher than 80\%. Third, we run this model on 10k projects, identifying and linking code summaries to the documented code. This unlocked the possibility of building a large-scale dataset of documented code snippets that have then been used to train a new DL model able to automatically document code snippets. A comparison with state-of-the-art baselines shows the superiority of the proposed approach, which however, is still far from representing an accurate solution for snippet summarization. %Still, our dataset, the classification \& linking model, and the large-scale dataset we built represent substantial steps ahead in the automated documentation of code snippets.
\end{abstract}




\maketitle

% !TEX root = main.tex
%%%%%%%%%%%%%%%%%%%%%%%%%%%%%%%%%%%%%%%%
%%%%%%%%%%%%%%%%%%%%%%%%%%%%%%%%%%%%%%%%
\section{Introduction} \label{sec:intro}
%%%%%%%%%%%%%%%%%%%%%%%%%%%%%%%%%%%%%%%%
%%%%%%%%%%%%%%%%%%%%%%%%%%%%%%%%%%%%%%%%

Empirical studies showed that code comprehension can take up to 70\% of developers' time \cite{Minelli:icpc2015,Xia:tse2018}. While code comments can support developers in such a process \cite{deSouza:2005}, their availability \cite{Spinellis:IE} and consistency with the documented code \cite{Fluri:wcre07,Fluri:SQJ09,Linares:ASE15} cannot be taken for granted. A helping hand may come from tools proposed in the literature to automatically document code \cite{Rahman:SCAM15,Rodeghero:icse17,Haiduc:wcre2010,sridhara2011automatically,wong2015clocom, wong2013autocomment,moreno2013automatic,sridhara2011automatically,McBurney:tse2016,iyer:acl,allamanis2016convolutional,aghaj:2019a,LeClair:icse2019,Hu:emse2020,haque:2020,Zhang:icse2020}. The most recent techniques (\eg \cite{LeClair:icse2019,Hu:emse2020,haque:2020}) train deep learning (DL) models with the aim of learning how to summarize a given piece of code in natural language. This requires the building of a large-scale dataset composed by pairs $<$$code$, $description$$>$ that can be used to feed the model with $code$ instances asking it to generate their $description$. These approaches are usually trained to work at function-level granularity. This means that, in the case of Java, methods are mined from open source projects and linked to the first sentence of their Javadoc which is assumed to represent a plausible code summary.

Having such a granularity could be, however, suboptimal to support comprehension activities. Indeed, while the overall goal of a method might be clear to a developer, they may not understand a specific set of statements in it. Also, looking at the datasets used in the literature to train these models, we found that the methods' descriptions extracted from the Javadoc are usually very short. For example, the seminal dataset by LeClair and McMillan \cite{leclair2019recommendations}, features an average of 7.6 words (median=8.0) to summarize each Java method. While such short descriptions could provide a grasp about the overall goal of the method, it is unlikely that they can actually support a developer struggling to understand it. 

For this reason, a few attempts have been made to automatically summarize code snippets rather than entire functions \cite{Rahman:SCAM15,wong2015clocom, wong2013autocomment,aghaj:2019a,sridhara2011automatically,wang2017automatically,huang2020towards}. Most of them are based on information retrieval \cite{Rahman:SCAM15,wong2015clocom,wong2013autocomment,aghaj:2019a} meaning that, given a code snippet $CS$ to document, the most similar snippet to it is identified in a previously built dataset and its comments are reused to summarize $CS$. These approaches, while valuable, rely on manually crafted heuristics to automatically identify the ``scope of an inner comment'', \ie the statements that a given comment documents. For example, one may assume that an \texttt{//inline comment} in Java documents all following statements until a blank line is found \cite{chen2019automatically}. As we will show, such a heuristic fails in several cases. Other techniques \cite{sridhara2011automatically,wang2017automatically} exploit pre-defined templates to document code snippets that, however, cannot generalize to all combinations of code statements one could find. 

Given the limitations of previous work, Huang \etal \cite{huang2020towards} proposed an approach exploiting reinforcement learning to document code snippets. The first challenge they faced was the creation of a training dataset. Indeed, while it is relatively easy to collect pairs of $<$$code$, $description$$>$ when working at function-level granularity, this is not the case for code snippets. For this reason, Huang \etal exploited an approach proposed by Chen \etal \cite{chen2019automatically} to automatically detect the scope of code comments. The approach exploits a combination of heuristics and learning-based techniques to automatically identify, given a comment, the set of statements documented by it. Using this approach, Huang \etal \cite{huang2020towards} built a dataset of $\sim$124k $<$$snippet$, $description$$>$ pairs which has been used to train \emph{RL-BlockCom}, a DL model combining reinforcement learning with a classic encoder-decoder model. \emph{RL-BlockCom} is able, given a code snippet as input, to automatically document it reaching a BLEU-4 \cite{papineni2002bleu} of 24.28. While being the first DL-based approach to support code snippets' summarization, \emph{RL-BlockCom} suffers of some major limitations mostly related to the way in which its training/test sets have been built exploiting the approach in \cite{chen2019automatically}:

\emph{1. Simplified/unrealistic linking of code comment to the documented snippet \cite{chen2019automatically}.} This is due to some of the design choices made in the scope detection approach \cite{chen2019automatically}. For example, the authors ``\emph{regard the first out-of-scope statement as the demarcation point of the scope of the comment}''. This means that, accordingly to their approach, it is not possible for a code comment to document non-contiguous statements. As we will show, our manual validation of 6,645 instances reveals 1598 ($\sim$27\%) cases of code comments that document non-contiguous statements. These are all cases which cannot be successfully supported by the scope detection approach and, as a consequence, by \emph{RL-BlockCom}.

\emph{2. Lack of filters to identify code summaries \cite{chen2019automatically}.} Chen \etal correctly observed that not all comments ``describe'' code statements. Thus, they use heuristics to remove commented out code, TODO comments, IDE-generated comments, and non-text comments containing dates or links. Despite these filters, using such an approach to create a training dataset for a snippet summarization approach such as \emph{RL-BlockCom} means feeding it with comments which may not be an actual code summary of the documented snippet. For example, when manually looking at the previously mentioned 6,645 instances, we found 33\% of them to just act as a logical split of source code (\ie a ``formatting'' comment \cite{pascarella2017classifying}) without providing additional information on the documented code (\eg a comment \texttt{//get messages} put on top of a method call \texttt{getMessages()}). These comments are useless to train a code summarizer, but are not excluded from the \emph{RL-BlockCom} training dataset.

\emph{3. The training dataset used in RL-BlockCom includes code summaries as short as two words \cite{huang2020towards}.} These are unlikely to be code summaries useful to support program comprehension.\vspace{0.05cm}

\begin{figure*}[tb]
    \centering
    \includegraphics[width=\linewidth]{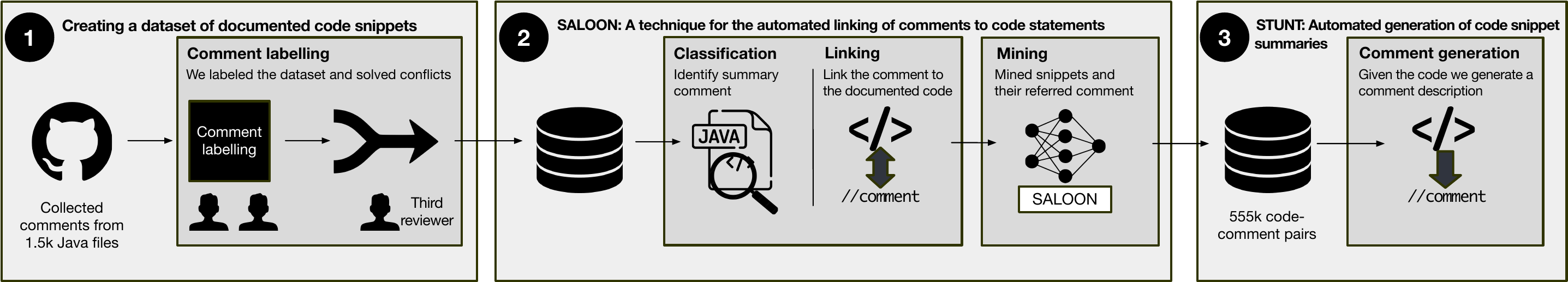}
    \caption{Approach Overview}
    \label{fig:specific}
\end{figure*}

To address these limitations, in this work we take all steps needed to foster the research on snippets summarization, as depicted in \figref{fig:specific}. First (step 1 in \figref{fig:specific}), we manually built a dataset of 6,645 $<$$snippet$, $description$$>$ pairs, in which we classified the code comment ($description$) as being or not a code summary and linked it to the documented Java statements. Such a dataset has been built by ensuring two evaluators for each analyzed comment, with a third one solving conflicts when needed. The overall effort spent by the six involved authors accounts for 815 man-hours. 

We use this dataset to fine-tune SALOON (step 2 in \figref{fig:specific}), a multi-task pre-trained Text-to-Text-Transfer-Transformer (T5) \cite{raffel2019exploring} model able to take as input an inner comment in a method and (i) classify whether it represents a valid code summary with a 83\% accuracy; and (ii) link it to the relevant code snippets it documents with a recall/precision higher than 80\%. We show that the performance of SALOON are significantly better than the comment-to-code linking approach by Chen \etal \cite{chen2019automatically}. 

Finally (step 3 in \figref{fig:specific}), we run SALOON on 10k GitHub Java projects to automatically build a large-scale dataset of $\sim$554k $<$$snippet$, $description$$>$ pairs. The latter has been used to train and test STUNT, a DL-based approach taking as input a code snippet and automatically generating its code summary. We show that STUNT performs better than IR-based and RL-based baselines \emph{RL-BlockCom}. 

Despite this finding, our results also show that STUNT is not yet ready to be deployed to developers and point to more research being needed on the task of snippet summarization.

In summary, our contributions are: (i) the largest manually built dataset in the literature featuring classified and linked code comments; (ii) SALOON, a multi-task DL model able to achieve state-of-the-art performance in the tasks of comment classification and linking; and (iii) STUNT, a code snippet summarization model trained on a large-scale and more realistic dataset as compared to the one used in the literature \cite{huang2020towards}. The dataset and all code used to train and test the models in this paper are available in our replication package \cite{replication}.

% !TEX root = main.tex
%%%%%%%%%%%%%%%%%%%%%%%%%%%%%%%%%%%%%%%%
%%%%%%%%%%%%%%%%%%%%%%%%%%%%%%%%%%%%%%%%
\section{Building a Dataset of Documented Code Snippets} \label{sec:dataset}
%%%%%%%%%%%%%%%%%%%%%%%%%%%%%%%%%%%%%%%%
%%%%%%%%%%%%%%%%%%%%%%%%%%%%%%%%%%%%%%%%

We detail the process used to build a manually validated dataset featuring triplets $<$$D$, $\{CC\}$, $DC$$>$ where $D$ represents a natural language comment documenting the code snippet $DC$ (\emph{Documented Code}) and $\{CC\}$ represents the \emph{Comment Category} (\eg code summary, TODO comment), with more than one category possibly being assigned to the comment. We later use such a dataset to train and evaluate the model described in \secref{sec:linking}, taking as input a comment $D$ and automatically (i) classifying it, thus being able to check whether $D$ is a code summary (\ie an actual description of the documented code) or another type of comment (\eg TODOs), and (ii) linking $D$ to the corresponding documented code $DC$. %Such a model has then be used to automatically build a large-scale dataset of documented code snippets $<$$D$, $DC$$>$, in which each $D$ is, accordingly to our model, a code summary. Such a dataset has then been used to train the model described in \secref{sec:generating} able to automatically document a code snippet provided as input.

\subsection{Study Design}
As a first step to build our dataset we needed to collect the set of code comments $D_1$, $D_2$, $\dots$, $D_n$ to manually analyze. To collect these comments, we used the web application by Dabic \etal~\cite{dabic2021sampling} to query GitHub for all Java projects having at least 500 commits, 25 contributors, 10 stars, and not being forks. These filters aim at discarding personal/toy projects and reducing the chance of mining duplicated code. The focus on Java was dictated by the will of accommodating the expertise of the manual validators (\ie the authors) all having extensive knowledge of the Java programming language. Despite the focus on Java, our methodology to build the dataset as well as to train the models described in the subsequent sections is general and can be reproduced for different languages.

%768,949

We randomly cloned 100 of the 1,681 projects resulting from our search on GitHub, for a total of $\sim$768k Java files. 

We parsed their code to identify comments within each method to manually analyze. We ignored Javadoc comments since they document entire methods rather than code snippets: We only considered single-line (starting with ``\texttt{//}'') and multi-line (starting with ``\texttt{/*}'') comments as subject of our manual analysis. Also, we did not extract comments from test methods (\ie methods annotated with \texttt{@Test}) to increase the cohesiveness of our dataset and only focus on documentation related to production code. The manual analysis has been performed by the six authors (from now on, evaluators) through a web app we developed to support the process.

%Finally, we ignored all comments featuring less than 5 tokens in an attempt to exclude comments unlikely to provide a description of the documented code. While we acknowledge that some very short comments could provide important insights about a code snippet, we preferred to focus our manual effort on longer instances more likely to represent interesting entries for our dataset. 

 We targeted the labeling of valid comments (\ie excluding those removed by the above-described procedure)  within 1,500 Java files, with the idea of creating a dataset of $\sim$10k triplets ($<$$D$, $\{CC\}$, $DC$$>$). The web app assigned each Java file to two evaluators who independently labeled the comments in it. If the number of comments in a file was higher than 10, the web app randomly selected a number of comments to label going from 10 to $m$, where $m$ was the actual number of valid comments in the file. Otherwise all comments in the file were labeled. We opted for this process to avoid an evaluator being stuck too much time on a single file. Also, we did not consider comments belonging to methods longer than 1,024 tokens and made sure no duplicated methods were present in the final dataset (\ie the same method might be present across different files/projects). The filter on the method length was driven by the final usage we envision for our dataset, namely training DL-models which usually works on inputs of limited size ($\leq$512 tokens, or even less, see \eg \cite{LiuCommit,Tufano:tosem2019, mastropaolo2022using,Tufano:icse2022,haque:2020,tufano2019learning}). Thus, labeling instances longer than 1,024 tokens would have been a waste of resources.

%\eject

The goal of the labeling was to firstly assign the comment $D$ to one or more categories $CC$s. The starting set of categories to use was taken from the work by Pascarella \etal \cite{pascarella2017classifying} and included: \emph{summary}, \emph{rationale}, \emph{deprecation}, \emph{usage}, \emph{exception}, \emph{TODO}, \emph{incomplete}, \emph{commented code}, \emph{formatter}, and \emph{pointer}. We do not describe these categories due to the lack of space, pointing the reader to \cite{pascarella2017classifying} for a complete description. However, as concrete examples, \emph{summary} represents the classic code description explaining what the code is about, \emph{formatter} is a comment used by developers to better organize the code into logical sections, while \emph{pointer} refers to comments linking external resources. We excluded from the original list by Pascarella \etal \cite{pascarella2017classifying} the following categories (i) \emph{directive} and \emph{autogenerated} since, as described by the authors, they both concern comments automatically generated by the IDE; and (ii) \emph{license} and \emph{ownership}, since this information is usually featured in Javadoc comments. 

Finally, we merged the \emph{expand} category into \emph{summary}, since the former is defined by the authors as a code description providing more information than a usual summary. Such a distinction is irrelevant for our work. Besides the set of predefined categories, we also gave the possibility to evaluators to define new categories. If an evaluator defined a new category, it was immediately visible to all other evaluators. The following additional categories have been defined by us: \emph{orphan}, indicating a code comment not linked to any line of code, and \emph{code example}, indicating a comment describing \eg how to invoke a specific method.

Once the category for a given comment under analysis was defined, the next step was the linking of the comment to the documented code $DC$. The linking has been performed at line-level granularity. This means, for example, that for a comment $D$ the evaluator could indicate lines 11, 12, and 17 as documented. Note that gaps are possible in $DC$ (\ie the documented code could be composed by non-contiguous lines). Our replication package \cite{replication} shows concrete examples of this scenario, that we omit here due to space limitations. Then, we started resolving conflicts arisen from the manual analysis. Two types of conflicts are possible for each manually defined triplet $<$$D$, $\{CC\}$, $DC$$>$: The two evaluators could have (i) selected a different set $\{CC\}$ when classifying the comment; and (ii) identified different sets of lines ($DC$) documented by the comment. Out of the 6,645  manually labeled comments, 1,395 (21\%) resulted in a conflict: 1,144 were due to different comment categories selected by the evaluators; 47 to differences in the selected $DC$; 204 concerned both the categories and the $DC$. Conflicts were solved by a third evaluator not involved in the labeling of the conflicting instance. 

Overall, we spent 815 man-hours on the labeling and conflict resolution, manually annotating 6,645 comments (with two evaluators for each of them) coming from 1,508 Java files and 85 software projects. We labeled a bit more than the target 1,500 since multiple evaluators were working in parallel without noticing that we hit our target. The obtained dataset, publicly available in our replication package \cite{replication}, is briefly described in the following.

\subsection{Dataset} \label{subsec:dataset}

\begin{table}[ht]
\centering
    \caption{Dataset output of manual labeling}
    \label{tab:dataset}
    {\small
    \begin{tabular}{@{}lrrrr@{}}
    \toprule
    \multirow{2}{*}{\textbf{Category}} & \multirow{2}{*}{\textbf{\#Instances}}  & \multicolumn{3}{c}{\textbf{Documented Statements}}\\ \cline{3-5} 
    & & \textbf{mean} & \textbf{median} & \textbf{sd}\\\midrule
    Summary & 3,841 & 3.40 & 3.0 & 2.70\\
    Formatting & 2,209 & 2.32 & 2.0 & 2.65\\
    Rationale & 983  & 3.04  & 2.0 & 2.74\\
    TODO & 258 & 0.46 & 0.0 & 1.16\\
    Commented Code & 184 & 0.00 & 0.0 & 0.00\\
     Pointer & 33 & 2.66 & 2.0 & 5.27\\
    Orphan & 29 & 0.00 & 0.0 & 0.00\\
    Code Example & 9 & 1.77 & 2.5 & 1.48\\
     Deprecation & 7 & 3.14 & 3.0 & 1.34\\
    Incomplete & 2 & 1.5 & 1.5 & 0.70\\\midrule
   
    \textbf{Overall} & \textbf{6,645} & \textbf{1.83} & \textbf{1.60} & \textbf{1.80}\\\bottomrule
    \end{tabular}
    }
\end{table}

\tabref{tab:dataset} summarizes the dataset obtained as output of our analysis. We excluded from the table the categories for which we did not find any instance (\eg \emph{exception} \cite{pascarella2017classifying}, likely to be more prevalent in Javadoc comments). Since a single comment can be associated to multiple categories (\eg \emph{summary} and \emph{rationale}), the sum of the ``\#Instances'' column does not add up to the total number of comments we manually classified (\ie 6,645). 

Besides reporting the categories to which the comments in our dataset belong, \tabref{tab:dataset} also shows descriptive statistics related to the number of statements documented by comments belonging to different categories. As expected, \emph{orphan} and \emph{commented code} comments are not linked to any code statement. More than 80\% of \emph{TODO} comments are also not linked to any statement, since in many cases todos are related to \eg feature that must be implemented. Similarly, the only two \emph{incomplete} comments we found both of them not linked to any code: These are partially written comments needing rework. 

The most frequent category is, as expected, the \emph{summary} one (3,841 instances) grouping comments summarizing one or more code statements (on average, 3.40 statements). Another popular category is ``\emph{formatting}'', with 2,209 instances. 

While one could expect no code linked to formatting comments, this is actually not the case since we used such a category also for comments not adding new information to the documented code but just acting as a logical split of the code (\eg a comment \texttt{//get messages} put on top of a method call \texttt{getMessages()}).

Finally, comments explaining the \emph{rationale} for implementation choices account for 983 instances. While we focus on the generation of code summaries, these instances often contains interesting information that are hard to automatically synthesize and could represent a seed for future research.

Interestingly, 1,598 of the comments in our dataset ($\sim$27\%) include ``gaps'' in the linked code. This means, for example, that a comment documents lines 11, 12, and 17 (but not lines 13-16) --- see \cite{replication} for concrete examples. This means that approaches to automatically link comment and code must take such a scenario into account. 
Motivated by these insights, we fill this gap by creating a novel method for classifying and linking code comments, as elucidated in Section \ref{sec:linking}.
% !TEX root = main.tex
%%%%%%%%%%%%%%%%%%%%%%%%%%%%%%%%%%%%%%%%
%%%%%%%%%%%%%%%%%%%%%%%%%%%%%%%%%%%%%%%%
\section{Automatic Classification of Code Comments and Linkage to Documented Code} \label{sec:linking}
%%%%%%%%%%%%%%%%%%%%%%%%%%%%%%%%%%%%%%%%
%%%%%%%%%%%%%%%%%%%%%%%%%%%%%%%%%%%%%%%%

We start by presenting SALOON (claSsification And Linking Of cOmmeNts), the approach we devised for the classification of code comments and their linking to the documented code (\secref{sub:approachLinking}). Then, we discuss the design of the study we run to assess its accuracy (\secref{sub:designLinking}) and the achieved results (\secref{sub:resultsLinking}). 

Once trained, SALOON can be run on hundreds of projects to build a large-scale dataset featuring classified and linked code comments. While we could just refer to SALOON as a ``T5 model trained for comment classification and linking'', we preferred to name it to simplify the reading when we introduce the other T5 model we train for the task of code summarization (\secref{sec:generating}).

\subsection{Approach Description}
\label{sub:approachLinking}
SALOON is built on top of T5, a DL transformer-based model \cite{raffel2019exploring}. T5 has been presented by Raffel \etal \cite{raffel2019exploring} as a model that can be trained to support any Natural Language Processing (NLP) task that can be represented in a text-to-text format, meaning that both the input and the output of the model are text strings. Such a representation is well-suited for code-related tasks, as demonstrated by the recent literature (see \eg \cite{wang-etal-2021-codet5,mastropaolo2021studying,Tufano:icse2022}). %Also, T5 is able to detect hidden and long-ranged dependencies among tokens, without assuming that nearest tokens are more related than distant ones. This last property is particularly relevant in code-related tasks and especially for the task we aim at supporting, in which the documented code may be distant from the comment describing it.

Raffel \etal \cite{raffel2019exploring} reported state-of-the-art results for several NLP benchmarks, especially when leveraging the ``pretrain-then-finetune'' paradigm: The model is first pre-trained on a large dataset with the goal of learning patterns about the underlying language of interest (\eg Java). Then, it is fine-tuned to learn a specific task of interest (\eg code summarization). The pre-training is performed using self-supervised pre-training objectives such as the \emph{masked language model}. 

The idea is to provide the model with input sentences (\eg Java methods) in which a percentage of randomly selected tokens has been masked, with the model in charge of guessing them. This prepares the model's weights for the fine-tuning in which tailored datasets are used to teach the model the specific task to support (\eg pairs of code and comments). The pre-training phase is particularly important when the dataset used for the fine-tuning is expensive to build (\ie it requires manual validation) and, as a consequence, is limited in size. This is the case for our work, since our fine-tuning is performed on the dataset described in \secref{sec:dataset}, in which comments have been categorized and linked to the relevant statements.

In SALOON, we exploit the T5\textsubscript{\textit{small}} architecture described by Raffel \etal \cite{raffel2019exploring}. Due to space constraints, we point the reader to the original paper for all architectural details. We describe how we built the pre-training and fine-tuning datasets for the tasks of comment classification and linking.

\subsubsection{Pre-training Dataset} \label{subsec:pretraining-linking}

We start from the Java CodeSearchNet dataset \cite{husain2019codesearchnet}, which features $\sim$1.6M Java methods, $\sim$499k of which including a Javadoc. Given the tasks we aim at supporting (\ie automatic classification of code comments and linking to the code they document), there are two ``target languages'' we aim to expose to T5 during pre-training: Java code and technical natural language in the form of code comments. CodeSearchNet features both of them.
 We preprocess the dataset by discarding all instances having \#tokens $>$ 1,024. During pre-training we treat Java methods and Javadoc comments as separated instances (\ie we ignore their association), thus removing Java methods and Javadoc comments being longer than 1,024 tokens. Such a filter removed $\sim$32k instances (\ie 31,702 methods and 178 Javadoc comments). Then, we excluded instances containing non-ASCII characters as well as Javadoc comments composed by less than 5 tokens (words), since unlikely to represent meaningful code descriptions ($\sim$57k instances removed). After removing duplicates, we end up with 1,870,888 pre-training instances (1,501,013 Java methods and 369,875 Javadoc).

\subsubsection{Fine-tuning Dataset} \label{subsec:ft-dataset-linking}

Two fine-tuning datasets are needed to support the tasks we target (\ie comment classification and linking). For comment classification, we built a dataset composed by pairs $\langle$$M_{j,D_i}$, $C_{c}$$\rangle$, in which a specific inner comment $D_i$ within a method $M_j$ is linked to a category $C_{c}$ classifying it (\eg code summary). For comment-to-code linking, we built a dataset featuring pairs $\langle$$M_{j,D_i}$, $DC$$\rangle$, in which $DC$ reports the $M_j$'s statements documented by $D_i$. Both datasets have been extracted from the manually built dataset of 6,645 classified and linked comments (\secref{sec:dataset}).

\textbf{Comment classification.} Given the goal of our work (\ie summarizing code snippets), we are interested in automatically identifying comments we classified as \emph{code summary} while excluding all the others. Starting from the dataset in \tabref{tab:dataset}, we extracted 3,841 $\langle$$M_{j,D_i}$, $C_{c}$$\rangle$ having $C_{c} = $ \emph{code summary} and 2,921 having having $C_{c} = $ \emph{other}. Basically, we target the training of a binary classifier taking as input a code comment ($D_i$) in the context of the method it belongs to ($M_j$) and guessing whether it is a code summary or not. 

The specific input we provide to T5 is $M_j$'s code with special tokens \texttt{<comment></comment>} surrounding the comment of interest (this is the representation of $M_{j,D_i}$), and expect as output either ``\emph{code summary}'' or ``\emph{other}'' (\ie $C_{c}$). 

Differently from the pre-training dataset, we did not need to remove sequences longer than 1,024 tokens, since this has already been done in the first place during the building of the dataset described in \secref{sec:dataset}. We randomly split the dataset into 80\% training, 10\% evaluation, and 10\% test. The first row in \tabref{tab:ft-dataset} shows the number of instances in these three sets.

\textbf{Code Linking.} Concerning the task of liking comments to code snippets, our training instances are only those comments that we manually labelled as \emph{code summary}. Indeed, we are interested in linking this specific type of comments to their code. Thus, we start from the 3,841 \emph{code summary} instances to build the needed $\langle$$M_{j,D_i}$, $DC$$\rangle$ pairs. Concerning the representation of $M_{j,D_i}$, it is similar to the previously discussed for the comment classification dataset (\ie the method $M_j$ with special tags surrounding the inner comment of interest $D_i$) with the only difference being a special tag \texttt{<N>} preceding each statement and reporting its line number in an incremental fashion.  

As for the expected output $DC$ (\ie documented code), it is represented as a stream of ``\texttt{<N>}'' tags representing the line numbers (\ie statements) within $M_j$ linked to $D_i$ (\eg \texttt{<1><2><4>}). Such a representation allows marking non-contiguous statements documented by $D_i$. 
The code linking fine-tuning dataset is composed by 3,841 instances split into 80\% training, 10\% evaluation, and 10\% test as shown in the second row of \tabref{tab:ft-dataset}. 
Note that to ensure a fair evaluation of the proposed approach, we split the dataset by taking into consideration the Java class from which these methods were originally extracted.

\begin{table}[h!]
	\centering
	\caption{Fine-tuning datasets}
	\small
	\begin{tabular}{lrrr}
		\toprule
		\textbf{Task}                  & \textbf{Train}    & \textbf{Eval}     & \textbf{Test}   \\
		\midrule
		\emph{Comment Classification}                   &  4,833       & 726     & 1,203        			               \\	
		\emph{Code Linking}                            & 2,805        &   403  &633                              \\
		%\midrule
		%\emph{\textbf{Total}}								& 13,342     & 1,887  &1,861				       \\
		\bottomrule
	\end{tabular}
	\label{tab:ft-dataset}
\end{table}

\subsubsection{Training Procedure and Hyperparameters Tuning} \label{subsec:hp-linking}
We evaluated the performance of eight T5 models (four pre-trained and four non pre-trained) on the evaluation set of each task in terms of correct predictions, namely cases in which the generated output (\ie the comment category or the documented statements) was identical to the expected output. 

We pre-train the T5 model from scratch (\ie starting from random weights) rather than starting from already pre-trained models for code such as CodeT5 \cite{wang2021codet5}, which is based on the same architecture proposed by Raffel \etal \cite{raffel2019exploring} we exploit in our investigation. Our decision is primarily motivated by the desire to have a model pre-trained on a single programming language (Java) as opposed to a multi-language model (as CodeT5).

We pre-train T5 for 300k steps using a 2x2 TPU topology (8 cores) from Google Colab with a batch size of 16. During pre-training, we randomly mask 15\% of tokens in an instance (\ie Java method or Javadoc comment), asking the model to guess the masked tokens. To avoid over-fitting, we monitored the loss function every 10k steps and stopped the training if such value did not improve after 12 consecutive evaluations (\ie after 120k steps, one epoch on our pre-training dataset). We use the  canonical T5$_{small}$ configuration \cite{raffel2019exploring} during pre-training. We also used the pre-training dataset to train a SentencePiece model (\ie a tokenizer for neural text processing) with vocabulary size set to 32k word pieces.

We fine-tuned a pre-trained and a non pre-trained model experimenting with four different learning rate schedulers (thus leading to eight overall trained models). 

Constant Learning Rate (C-LR) fixes the learning rate during the whole training; Inverse Square Root Learning Rate (ISR-LR), in which the learning rate decays as the inverse square root of the training step; Slanted Triangular Learning Rate (ST-LR), in which the learning rate first linearly increases and then linearly decays to the starting value; and Polynomial Decay Learning Rate (PD-LR), having the learning rate decaying polynomially from an initial value to an ending value in the given decay steps. The parameters used for the learning rates are available in \cite{replication}. 

\eject

We fine-tuned each of the eight models for a total of 75k steps on the fine-tuning training set of each task. We include in our replication package \cite{replication} a table showing the percentage of correct predictions (for the \emph{comment classification} task), precision and recall (for the \emph{code linking} task) achieved by each of the pre-trained and non pre-trained models on the evaluation sets. 

Overall, the pre-trained models work substantially better, especially when it comes to the \emph{code linking} task. In particular, in their respective best configuration, pre-trained models achieve (i) a 75\% classification accuracy in the \emph{comment classification} task as compared to the 58\% of the non pre-trained models; and (ii) 85\% precision and 89\% recall in the \emph{code linking} task, as compared to the 53\% precision and 67\% recall of the non pre-trained models. Such a result is expected considering that the fine-tuning training datasets are quite small due to the substantial manual effort required to build them ($\sim$6.7k instances for \emph{comment classification} and $\sim$3.8k for \emph{code linking}). Having small fine-tuning datasets is the scenario in which pre-training is known to bring major benefits \cite{Robbes:icse2019}.  As for the learning rate, the best results are achieved with ISR-LR when pre-training and with PD-LR when not pre-training. 

%the two different T5 models (\ie pre-trained and non pre-trained) we opted to focus only on the best-performing model (\ie T5 pre-trained) using the C-LR scheduler to fine-tune the final model in a multi-task fashion: this means creating a single T5 model that can support both tasks (\ie \emph{comment classification} and \emph{code linking}), thus leveraging a multi-task setting. In doing so, the model learns how to solve both tasks simultaneously, potentially reusing part of the knowledge gained to solve one task also to support the other task \cite{raffel2019exploring}}. 

To obtain the final model to use in SALOON, we fine-tuned the best performing model (\ie pre-trained with ISR-LR) using an early-stopping strategy in which we evaluated the model on the evaluation sets every 5k steps, stopping when no improvements were observed for 5 consecutive evaluations. We discuss the results achieved by SALOON as compared to other baselines in \secref{sub:resultsLinking}.

%\begin{table}[h!]
%	\scriptsize
%	 \resizebox{\columnwidth}{!}{
%	\begin{tabular}{lrr}
%		\toprule
%		\multicolumn{1}{c}{\textbf{Model}} & \multicolumn{2}{c}{\textbf{Task}} \\\cline{2-3}
%		& \textbf{Classification of Code Comments} & \textbf{Linking of Code Comments}  \\\midrule
%
%		\textbf{T5 Pre-Trained} & \textbf{82.60\%} & \textbf{45.17\%} \\
%		\textbf{T5 No Pre-Trained} & 79.00\% & 18.60\% \\
%	
%	
%		\bottomrule
%	\end{tabular}
%}
%	\vspace{0.3cm}
%	\caption{Achieved perfect predictions by the T5 pre-trained and non pre-trained models when supporting the tasks of automatic classification of code comments and linkage to documented code}
%	\label{tab:ablation}
%	
%\end{table}

\subsection{Study Design}
\label{sub:designLinking}

The goal of the study is to assess the accuracy of SALOON in the two tasks it has been trained for: \emph{comment classification} and \emph{code linking}. The context is represented by the test sets reported in \tabref{tab:ft-dataset}, featuring 1,203 instances for the task of comment classification and 633 for the task of code linking. 

Concerning the comment classification task, we do not compare SALOON against any baseline, since our goal (\ie identifying only code summaries) is quite specific of our work. Instead, we compare the performance of SALOON against the three following baselines for the task of code linking (the implementation of all baselines is publicly available \cite{replication}).

\textbf{Heuristic-1: blank line \cite{chen2019automatically}.} The first baseline is a straightforward heuristic assuming that a given \texttt{//inline comment} documents all following statements until a blank line is reached.

\textbf{Heuristic-2: token-based string similarity \cite{Fluri:wcre07}.} The basic idea of this heuristic is that statements sharing terms with a code comment are more likely to be documented by it. We use the token-based string similarity by Fluri \etal \cite{Fluri:wcre07} to compute the textual similarity between each comment in the test set and all statements in the method it belongs to. A statement is linked to the comment if its similarity with it is higher or equal than a threshold $\lambda$. The similarity is computed as the percentage of overlapping terms between the two strings (\ie comment and statement), with the terms being extracted through space splitting. We experiment with different values for $\lambda$, going from 0.1 (\ie 10\% of terms are shared between the two strings) to 0.9 at steps of 0.1. 

\eject

\textbf{ML-based solution \cite{chen2019automatically}.} The approach by Chen \etal~\cite{chen2019automatically} relies on the random forest machine learning algorithm to classify statements in a method as linked or not to a given comment. Unfortunately, the source code of such approach is not available and, thus, we had to reimplement it following the description in the corresponding article. In a nutshell, the approach works as follows. The random forest uses three families of features to characterize a given statement and classify it as linked or not to a given comment. The first family comprises eight ``code features'', capturing characteristics of the statement, such as the statement type (\eg \texttt{if}, \texttt{for}) and whether the statement shares method calls with the statements preceding and following it. The second family includes four ``comment features'', focusing on characteristics of the comment of interest, such as its length and the number of verbs/nouns it contains. Finally, the third family groups four ``relationship features'', representing the relationship between the comment and the statement (\eg textual similarity). For a fair comparison, we train the random forest on the same training set used for SALOON.

\subsubsection{Data Collection And Analysis} \label{sub:linking-analysis}
Concerning the \emph{comment classification} task, we run SALOON on the test set and report the accuracy of the model in classifying comments representing ``code summaries''. As for the \emph{code linking}, we start computing the percentage of \textbf{correct predictions}, namely cases in which all statements linked to a comment in the test set match the ones in the oracle. This means that a comment instance correctly linked to two out of the three statements it documents is considered wrong. We also compute the \textbf{recall} and \textbf{precision} of the techniques at statement-level.  The recall is computed as TP/(TP+FN), where TP represents the set of code-to-comment links correctly identified by a technique (\ie a statement correctly linked to a comment) and FN are the set of correct code-to-comment links in the oracle missed by the approach. The precision is instead computed as TP/(TP+FP), with FP representing the code-to-comment links wrongly reported by the approach (\ie statements wrongly identified as linked to the comment). We also statistically compare the techniques assuming a significance level of 95\%. We compare  precision and recall using the Wilcoxon signed-rank test \cite{wilcoxon}. To control for multiple pairwise comparisons (\eg SALOON's precision compared with that of the three baselines), we adjust $p$-values with Holm's correction \cite{Holm1979a}. 

We estimate the magnitude of the differences using the Cliff's Delta ($d$), a non-parametric effect size measure \cite{Gris2005a}. We follow well-established guidelines to interpret the effect size: negligible for $|d| < 0.10$, small for $0.10 \le |d| < 0.33$, medium for $0.33 \le |d| < 0.474$, and large for $|d| \ge 0.474$ \cite{Gris2005a}. As for the percentage of correct predictions, we pairwise compare them among the experimented techniques, using the McNemar's test \cite{mcnemar}, which is a proportion test suitable to pairwise compare dichotomous results of two different treatments. We complement the McNemar's test with the Odds Ratio (OR) effect size. Also in this case we use the Holm's correction procedure \cite{Holm1979a} to account for multiple comparisons.

%Finally, only for T5, we report the percentage of \textbf{broken predictions}. To explain those cases let us remind that T5 takes as input a method with the comment of interest surrounded by special tags and generates as output the same method with injected $\langle$$start$$\rangle$ and $\langle$$end$$\rangle$ tokens surrounding the candidate documented statements. However, it could happen that (i) in this process, T5 alters the code of the input method despite the fact that it should simply copy it and inject the $\langle$$start$$\rangle$ and $\langle$$end$$\rangle$ tokens; or (ii) the injected $\langle$$start$$\rangle$ and $\langle$$end$$\rangle$ tokens are not correctly balanced (\eg it injects an extra $\langle$$end$$\rangle$ token). These are cases we treat as ``broken predictions''.

\subsection{Results Discussion}
\label{sub:resultsLinking}

As for the  \textit{comment classification} task, SALOON correctly classifies 78.05\% (939/1,203) of instances. Out of the 633 \emph{code summary} comments present in the test set, 536 (84\%) have been correctly classified, while 97 have been mistakenly reported as \emph{other}. 

Concerning the 570 ``\emph{other}'' comments, SALOON correctly predicted 403 (70\%) of them, wrongly reporting 167 instances as \emph{code summary}. This results in a recall=0.85 and precision=0.76 when identifying a comment as a \emph{code summary}. This means that by running our approach on the comments of a previously unseen software system, we can expect to identify 85\% of code summaries present in it accompanied, however, by 25\% of false positives (\ie non \emph{code summary} comments). 

\begin{table}[h!]
	\centering
	\caption{T5 \emph{vs} baselines on the code linking task\vspace{-0.2cm}}
	\scriptsize
	\begin{tabular}{lrrr}
		\toprule
		\textbf{Technique} & \textbf{Correct Predictions}                  & \textbf{Recall}    & \textbf{Precision} \\
		\midrule
		\emph{Blank line \cite{chen2019automatically}} & 0.20                   &  0.87       & 0.57          			               \\	\midrule
		\emph{Token-based similarity \cite{Fluri:wcre07}}                            &         &                                 \\
		\hspace{0.2cm} $\lambda$=0.1 & 0.03 & 0.62 & 0.33\\
		\hspace{0.2cm} $\lambda$=0.2 & 0.05 & 0.38 & 0.34\\
		\hspace{0.2cm} $\lambda$=0.3 & 0.05 & 0.23 & 0.26\\\midrule
		%\hspace{0.2cm} $\lambda$=0.4 & 0.03 & 0.09 & 0.53\\
		%\hspace{0.2cm} $\lambda$=0.5 & 0.02 & 0.03 & 0.57\\\midrule
		\emph{ML-based \cite{chen2019automatically}} & 0.23 & 0.49 & 0.58\\\midrule
		\emph{SALOON} & \textbf{0.58} & \textbf{0.89} & \textbf{0.86}\\\bottomrule
	\end{tabular}
	\label{tab:linking-results}
\end{table}

Concerning the \textit{code linking} task, \tabref{tab:linking-results} reports the correct predictions (\ie for a given comment in our test set \textbf{all} linked statements have been correctly identified), recall, and precision achieved by SALOON and the three baselines. \tabref{tab:stats-linking} reports the results of the statistical tests. For the Cliff's Delta $d$ we use N, S, M, and L to indicate its magnitude from Negligible to Large.

Note that for the \emph{token-based string similarity} baseline we report the results achieved with different values of $\lambda$ (\ie minimum similarity threshold to link a code statement to a comment). 

While we also experimented with values going up to 0.9 \cite{replication}, the recall values were too close to 0 to consider these variants as reasonable baselines. %Indeed, as expected, increasing the value of $\lambda$ results in an increase of precision accompanied, however, by a decrease of recall.

%\begin{table}[h!]
%\centering
%		\caption{Statistical tests: SALOON \emph{vs} baselines.}
%		\tiny
%		\label{tab:tests}
%		\begin{tabular}{lrrlrrlrr}
%			\toprule
%			\multirow{2}{*}{\textbf{Comparison}}  
%				& \multicolumn{2}{c}{Correct Predictions} &
%				& \multicolumn{2}{c}{Recall} &
%				& \multicolumn{2}{c}{Precision}\\\cline{2-3} \cline{5-6} \cline{8-9}
%			& \textbf{\emph{p}-value} & \textbf{\emph{OR}} &
%			& \textbf{\emph{p}-value} & \textbf{\emph{d}} &
%			& \textbf{\emph{p}-value} & \textbf{\emph{d}} \\
%			\midrule
%			SALOON \emph{vs} blank line  \cite{chen2019automatically} & $<$0.001 & 10.44 && 0.87 &  -0.00 (N) && $<$0.001 & -0.39 (M)\\
%			 SALOON \emph{vs} token sim.(0.1) \cite{Fluri:wcre07}  & $<$0.001 & 42.80 && $<$0.001 & -0.46 (M) && $<$0.001 & -0.67 (L)\\
%			SALOON \emph{vs}  token sim.(0.2) \cite{Fluri:wcre07} & $<$0.001 & 24.18 && $<$0.001 & -0.67 (L) && $<$0.001 & -0.54 (L)\\
%			 SALOON \emph{vs} token sim. 0.3) \cite{Fluri:wcre07} & $<$0.001 & 24.47 && $<$0.001 & -0.79 (L) && $<$0.001 & -0.49 (L)\\
%			 SALOON \emph{vs} ML-Based \cite{chen2019automatically} & $<$0.001 & 10.88 && $<$0.001 & -0.51 (L) && $<$0.001 & -0.17 (S)\\
%			\bottomrule
%		\end{tabular}
%
%\label{tab:stats-linking}
%\end{table}

\begin{table}[ht]
	\centering
	\caption{Code linking task: SALOON \emph{vs} baselines\vspace{-0.2cm}}
	\label{tab:stats-linking}
	\begin{adjustbox}{width=\columnwidth,center}
	\begin{tabular}{llrcc}
		\hline
		\textbf{Comparison} & \textbf{Metric} & \textbf{\emph{p}-value} & \textbf{d} & \textbf{OR}\\ 
		\hline
		\multirow{3}{*}{Blank line \cite{chen2019automatically} \emph{vs} SALOON} & Correct Predictions & $<$0.05 & - & 19.28\\ 
		&Recall & $<$0.05 & -0.04 (N) & - \\ 
		& Precision & $<$0.05& -0.48 (L) & -\\\midrule
		
		\multirow{3}{*}{Token sim.(0.1) \cite{Fluri:wcre07} \emph{vs} SALOON} & Correct Predictions & $<$0.05 &- & 70.80\\ 
		&Recall & $<$0.05& -0.45 (M) & - \\ 
		& Precision & $<$0.05 & -0.75 (L) & -\\\midrule
		
		\multirow{3}{*}{Token sim.(0.2) \cite{Fluri:wcre07} \emph{vs} SALOON} & Correct Predictions & $<$0.05 & - & 37.77\\ 
		&Recall & $<$0.05 & -0.66 (L) & -\\ 
		& Precision & $<$0.05& -0.68 (L) & -\\\midrule
		
		\multirow{3}{*}{Token sim.(0.3) \cite{Fluri:wcre07} \emph{vs} SALOON} & Correct Predictions & $<$0.05 & - & 38.00\\ 
		&Recall & $<$0.05 & -0.80( L) & -\\ 
		& Precision & $<$0.05 &  -0.73 (L)& -\\\midrule
		
		\multirow{3}{*}{ML-Based \cite{chen2019automatically} \emph{vs} SALOON}& Correct Predictions & $<$0.05 & - & 15.80\\ 
		&Recall & $<$0.05 & -0.49 (L) & -\\ 
		& Precision & $<$0.05 & -0.33 (M) & -\\
		\hline
	\end{tabular}
\end{adjustbox}
\vspace{-0.2cm}
\end{table}

SALOON predicts all statements linked to a given comment in 58\% of cases, against the 23\% achieved by the best-performing baseline (\emph{ML-based}). The \emph{blank-line technique} achieves 20\% of correct predictions.

The results of the statistical tests confirm the better performance ensured by SALOON in terms of correct predictions: McNemar's test always indicates significant differences in terms of correct predictions accompanied by ORs indicating that SALOON has between 15.80 to 70.80 higher odds of providing a correct prediction against the baselines.

Recall and precision values confirm the superiority of SALOON for the \emph{code linking} task. In terms of recall, SALOON is able to correctly link 89\% of statements in our dataset, achieving the best performance among all the experimented techniques. While the \emph{blank-line} approach achieves a similar recall (87\%) it pays a much higher price in terms of precision, with a 43\% false positive rates as compared to the 14\% of SALOON. Note that a high recall for this heuristic is expected, considering that it links all statements following a comment until a blank line is found. The \emph{ML-Based} technique can only predict half of the correct links (0.49) while achieving a precision score of 0.58. Accordingly to our results, the \emph{token-based similarity} heuristic does not represent a viable solution for the \emph{code linking} task: The best results are achieved when considering ($\lambda$=0.1) as a threshold, for which the technique can ensure a recall of 0.62 and a precision of 0.33. Differences in terms of recall and precision are always statistical significant (see \tabref{tab:stats-linking}). The effect size is in most of cases medium or large, with the only exception of the recall test comparing T5 with the \emph{blank-line} baseline, for which a negligible effect size is reported.

To summarize, SALOON is able to identify comments representing code summaries with a recall of 0.85 and a precision of 0.76. Also, it achieves state-of-the-art results in linking comments to the documented code, with a recall of 0.89 and a precision of 0.86. In \secref{sec:generating} we explain how we exploit this model to build a large-scale dataset aimed at training a T5 fine-tuned for the task of code snippet summarization.
% !TEX root = main.tex
%%%%%%%%%%%%%%%%%%%%%%%%%%%%%%%%%%%%%%%%
%%%%%%%%%%%%%%%%%%%%%%%%%%%%%%%%%%%%%%%%
\section{Snippets Summarization Using T5} \label{sec:generating}
%%%%%%%%%%%%%%%%%%%%%%%%%%%%%%%%%%%%%%%%
%%%%%%%%%%%%%%%%%%%%%%%%%%%%%%%%%%%%%%%%

We discuss how we trained a T5 model for the task of code snippet summarization (\secref{sub:approach_generation}), the study we run to evaluate it (\secref{sub:design_generation}) and the achieved results (\secref{sub:results_generation}). We refer to the snippet summarization approach as ``STUNT'' (SnippeT sUmmarizatioN using T5).

\subsection{Approach Description} \label{sub:approach_generation}
We rely on the same T5 architecture described in \secref{sub:approachLinking} and we reuse the same pre-trained model we built for the \emph{comment classification} and \emph{code linking} tasks. Indeed, as explained in \secref{subsec:pretraining-linking}, we pre-trained the model on a dataset composed by $\sim$1.5M Java methods and their inner comments and $\sim$370k Javadoc comments. Thus, T5 has been pre-trained to acquire knowledge about the two ``target languages'' relevant for the summarization task as well (\ie Java code and technical language used to summarize it). We detail the fine-tuning dataset and the training procedure.

\subsubsection{Fine-tuning Dataset}\label{subsec:ft-dataset-summarizer}
We used the GHS tool by Dabic \etal~\cite{dabic2021sampling} to query GitHub for all public non-forked Java projects with minimum 50 commits, 5 contributors, and  10 stars. The idea of these filters was to remove toy/personal projects while still obtaining a large set of projects to provide as input to SALOON with the goal of identifying comments representing summaries and linking them to the relevant code. We cloned 10k of the 18.7k projects returned by our query and extracted their methods using srcML \cite{collard2013srcml}. 

We excluded all methods longer than 512 tokens and removed all duplicates, obtaining a set of methods $S$. We also removed duplicates between our pre-training dataset and $S$ and between our manually labeled dataset (\secref{subsec:dataset}) and $S$. 

Concerning the removal of duplicates between the pre-training dataset and $S$, this was needed since $S$ is our starting point to build the fine-tuning dataset for the snippet summarization task from which we will also extract the test set on which STUNT will be evaluated. Thus, we ensure that STUNT is not evaluated on already seen instances. As for the removal of duplicates between the manually labeled dataset and $S$, this is due to the fact that SALOON (\ie our approach for comment classification and linking) has been trained on those instances and we will run it on $S$ to build the fine-tuning dataset for STUNT (\ie for code summarization). Running SALOON on already seen instances would inflate its performance, and not provide a realistic picture of what can be achieved by training STUNT on a dataset automatically built using SALOON.

From the remaining methods, we extracted all inner comments, filtering out those shorter than 5 words (unlikely to represent a meaningful code summary). As done in previous code summarization works \cite{LeClair:icse2019}, we lowercased and stemmed the comments (using the spaCy NLP library \cite{spacy}). Then, for each comment $D_i$ extracted from a method $M_j$ we created an instance $M_{j,D_i}$ in which $M_j$'s code features special tokens \texttt{<comment></comment>} to surround the comment of interest ($D_i$). This means that if $M_j$ features three inner comments, three $M_{j,D_i}$ instances will be created, each having a different comment ($D_i$) ``tagged''. This format is the one expected by SALOON to automatically (i) classify $D_i$ as \emph{code summary} or \emph{other}, and (ii) link $D_i$ to the relevant code statements. 

The above-described process resulted in 2,210,602 $M_{j,D_i}$ instances that we provided as input to SALOON, which classified 907,660 of them as \emph{code summary}. Among these, SALOON automatically linked code statements to the \emph{code summaries} in $\sim$85\% of cases (776,531). These instances are $\langle$$M_{j,DC}$, $D_i$$\rangle$ pairs, where $M_{j,DC}$ represents the method $M_j$ with special tokens \texttt{<start><end>} surrounding the statements ($DC$) documented by $D_i$. 

If more non-contiguous statements are documented, multiple \texttt{<start><end>} pairs are injected in $M_j$. These pairs are those needed to fine-tune STUNT for the task of snippet summarization: the input provided to the model is $M_{j,DC}$ (\ie a snippet to document) and the expected output is the documentation $D_i$. To avoid favoring the model during testing, we also removed all duplicates at snippet-level granularity. This means that if we have in our dataset two different methods containing the same $DC$ (\ie the same code snippet to document), we only keep one of them. Also, being SALOON an automated approach, it is expected to produce wrong instances (\eg comments linked to wrong statements) which, in turn, will penalize the performance of STUNT. By manually inspecting a sample of the pairs in our dataset, we noticed that one clear case of wrong instances are those in which the model had very low confidence in identifying the documented statements thus producing random symbols rather than the expected documented line numbers.  We automatically remove those instances, obtaining a set of 554,748 pairs, split into 80\% training (443,798), 10\% evaluation (55,475), and 10\% testing (55,475).

\subsubsection{Training Procedure and Hyperparameters Tuning}

As explained, we started from the already pre-trained T5 model. We then followed the same hyperparameters tuning discussed in \secref{subsec:hp-linking}, assessing the performance of four different learning scheduler on the evaluation set using the BLEU-4 score \cite{papineni2002bleu} as performance metric. The BLEU-4 variant computes the BLEU score by considering the overlap of 4-grams between the generated text (\ie the synthesized snippet summary) and the target text (\ie the summary written by the original developers). This metric has been used by most of the previous work on code summarization (see \eg \cite{haque:2020, ahmad2020transformer, 9031440, leclair2020improved, 9609119, 8811932, Hu:icpc2018, huang2020towards, Hu:emse2020, hu2018summarizing, iyer2016summarizing, wan2018improving, wang2021cocosum, wei2020retrieve, ye2020leveraging, zhang2020retrieval}). Each of the four models has been trained for 100k steps before its evaluation. C-LR (\ie constant learning rate) provided the best performance. Data about this evaluation are available in our replication package \cite{replication}.

Once identified the best T5 variant, we fine-tuned it for up to 500k steps, using an early-stopping strategy to tame over-fitting. To this aim, we monitored the BLEU-4 score achieved on the evaluation set every 5k steps, stopping the training when no improvements were observed after 5 consecutive evaluations. %The model converged after 125k steps.

\subsection{Study Design} \label{sub:design_generation}

The goal is to assess the accuracy of STUNT for snippet summarization. The context is represented by (i) 55,475 $\langle$$M_{j,DC}$, $D_i$$\rangle$ pairs identified by SALOON as described in \secref{subsec:ft-dataset-summarizer} and belonging to the test set, and (ii) the test set made publicly available by Huang \etal \cite{huang2020towards} when presenting \emph{RL-BlockCom}, the state-of-the-art snippet summarization approach discussed in \secref{sec:intro}. 

We assess the performance of STUNT against an information retrieval (IR)-based technique (\ie IR-Jaccard) and \emph{RL-BlockCom}. To explain the basic idea behind the IR-based baseline let us remind that both our training and test set are composed by $\langle$$M_{j,DC}$, $D_i$$\rangle$ pairs. Given a pair in the test set, the baseline retrieves in the training set the pair having the $DC$ snippet being the most similar to the one in the test set pair. This means that this pair contains a documented snippet that is very similar to the one in the test set for which we have to generate a code summary. Once identified the most similar snippet in the training set, the IR-based technique reuses its description to document the instance in the test set. This baseline serves as a representative of works using IR to retrieve similar comments from a given dataset, including \eg \cite{wong2013autocomment}.

\textbf{IR: Jaccard index \cite{hancock2004jaccard}.} IR-Jaccard identifies the most similar snippet using the Jaccard similarity index. The latter considers the overlapping between two sets of unique elements, representing in our case the tokens composing the documented code ($DC$) in the test instance and in each of the training instances. Indeed, we need to compare each instance in the test set to all those in the training set to find the most similar one. The similarity is computed as the percentage of overlapping tokens between the two sets.

%\textbf{IR-2: TF-IDF \cite{ramos2003using}.} IR-2 exploits a Vector Space Model \cite{salton1975vector} with tf-idf \cite{ramos2003using} weighting schema to compute the similarity between the $DC$ in the test instance and in each of the training instances. Differently from the Jaccard index, repeated tokens in the compared strings influence the similarity value. \smallskip

An additional baseline for STUNT is \emph{RL-BlockCom} by Huang \etal \cite{huang2020towards}. Despite the code being available, we did not manage to re-train their approach on our dataset. We contacted the authors asking for help without, however, receiving answer. Thus, as an alternative form of comparison, we thought about training and testing STUNT on their dataset, which is publicly available, and then comparing the summaries generated by STUNT with those generated by \emph{RL-BlockCom}. Unfortunately, the authors did not make the summaries generated by their approach publicly available. The only viable form of comparison we found was to (i) re-train STUNT on the training dataset made available by Huang \etal \cite{huang2020towards} and used to train \emph{RL-BlockCom}; (ii) use this trained version of STUNT to generate predictions on the same test set on which \emph{RL-BlockCom} has been evaluated; (iii) use the evaluation scripts made available by Huang \etal for the computation of the sentence-level BLEU score; and (iv) compare the achieved results with those reported in their paper. Indeed, not having access to the summaries generated by \emph{RL-BlockCom} does not allow us to double-check the data reported in the original paper nor to compute additional metrics besides those used by the authors (BLEU). Note also that the training/test datasets shared by Huang \etal feature pairs $\langle$$DC$, $D_i$$\rangle$ as compared to our $\langle$$M_{j,DC}$, $D_i$$\rangle$ pairs. This means that STUNT cannot exploit the contextual information of the method $M_j$ when generating the predictions on their dataset.

\subsubsection{Data Collection And Analysis}
To compare the performance of our model against the two IR-based baselines, we exploit three metrics explained in the following. 

Out of those, only BLEU has been used in the comparison with \emph{RL-BlockCom} for the reasons previously explained.

\textbf{BLEU}~\cite{papineni2002bleu} assesses the quality of the automatically generated summaries by assigning a score between 0 and 1. In our case, 1 indicates that the natural language summary automatically generated is identical to the one originally written by the developer. Since in the test set we built there are no summaries shorter than 4 words, we use the BLEU-4 variant in the comparison with the IR-based baselines.  When comparing with \emph{RL-BlockCom} on their test set, we also compute BLEU-1, BLEU-2 and BLEU-3 as done by Huang \etal \cite{huang2020towards}.

\textbf{METEOR}~\cite{meteor} is a metric based on the harmonic mean of unigram precision and recall (the recall is weighted higher than the precision). Compared to BLEU, METEOR uses stemming and synonyms matching to better match the human perception of sentences with similar meanings. Values range from 0 to 1, with 1 being a perfect match.

\textbf{ROUGE}~\cite{lin2004rouge} is a set of metrics focusing on automatic summarization tasks. We use the ROUGE-LCS (Longest Common Subsequence) variant, which identifies longest co-occurring in sequence n-grams. ROUGE-LCS returns three values, the recall computed as \textit{LCS(X,Y)/length(X)}, the precision computed as \textit{LCS(X,Y)/length(Y)}, and the F-measure computed as the harmonic mean of recall and precision where \textit{X} and \textit{Y} represent two sequences of tokens.

We also statistically compare the different approaches assuming a significance level of 95\%. Also in this case we use the Wilcoxon signed-rank test \cite{wilcoxon}, adjusting $p$-values to account for multiple comparisons (Holm's correction procedure \cite{Holm1979a}) and the  Cliff's Delta ($d$) as effect size measure \cite{Gris2005a}. The statistical comparison was not possible with \emph{RL-BlockCom} since we only had access to the overall BLEU scores reported in the paper (\ie the  BLEU scores for each generated summary were not available).

\subsection{Results}\label{sub:results_generation}

\tabref{tab:rlcomBleu} compares STUNT and \emph{RL-BlockCom}, using the values reported in the paper by Huang \etal \cite{huang2020towards} as BLEU scores for \emph{RL-BlockCom}. STUNT achieves better performance for all BLEU scores, outperforming the state-of-the-art approach by a large margin (\eg +7 points of BLEU-4). A deeper comparison of the two techniques is not possible since the summaries generated by \emph{RL-BlockCom} are not available.

\begin{table}[h]
	\centering
	\caption{BLEU scores: STUNT \emph{vs} RL-BlockCom \cite{huang2020towards}}
	\label{tab:rlcomBleu}
	\small
	\begin{tabular}{lrrrr}
		\toprule
		%\multicolumn{4}{c}{{\bf RL-BlockCom \cite{huang2020towards}}}\\\midrule
		& {\bf RL-Com} && {\bf STUNT}\\\midrule
		BLEU-1 & 32.18  && \bf 34.17 \\
		BLEU-2 & 25.98  && \bf 31.09 \\
		BLEU-3 & 24.36  && \bf 30.63 \\
		BLEU-4 & 24.28  && \bf 31.22 \\\bottomrule	
	\end{tabular} 
\end{table}

\tabref{tab:large-scale-results} compares STUNT against IR-Jaccard on the large-scale dataset we built. Accordingly to all metrics used in our evaluation, the gap in performance between STUNT and the baseline (\ie IR-Jaccard) is substantial, with at least a +11 in terms of BLEU-4, a +12 in terms of ROUGE-LCS f-measure, and a +16 in terms of METEOR score. As observed by Roy \etal \cite{metricsImprovement}, METEOR is ``\emph{extremely reliable for differences greater than 2 points}'' in assessing code summarization quality as perceived by humans (\ie also humans are likely to prefer STUNT's summaries over those generated by the baselines).

%\begin{table}[h]
%	\centering
%	\caption{Evaluation Metrics: STUNT \emph{vs} IR-based baselines\vspace{-0.2cm}}
%	\scriptsize
%	\label{tab:large-scale-results}
%	\begin{tabular}{lrrrrr}
%		\toprule
%		& {\bf IR-1 (Jaccard)} & {\bf IR-2 (TF-IDF)} &  {\bf STUNT}\\\midrule
%		BLEU-4 \cite{papineni2002bleu}& 15.25 & 11.80  & \bf 31.40\\
%		ROUGE-LCS \cite{lin2004rouge} &  &  &\\
%		\hspace{0.2cm} $precision$ & 10.47 & 6.67 & \bf 25.87\\
%		\hspace{0.2cm} $recall$ & 10.57 & 6.93 & \bf 28.83\\
%		\hspace{0.2cm} $fmeasure$ & 9.64 & 6.05 & \bf 26.12\\
%		METEOR \cite{meteor} & 12.36 & 8.50  & \bf 33.10 \\\bottomrule
%	\end{tabular} 
%	\vspace{-0.2cm}
%\end{table}

\begin{table}[h]
	\centering
	\caption{Evaluation Metrics: STUNT \emph{vs} IR-Jaccard}
	\small
	\label{tab:large-scale-results}
	\begin{tabular}{lrrrr}
		\toprule
		& {\bf IR-Jaccard} &  {\bf STUNT}\\\midrule
		BLEU-4 \cite{papineni2002bleu}& 27.43   & \bf 38.42\\
		ROUGE-LCS \cite{lin2004rouge} &    &\\
		\hspace{0.2cm} $precision$ & 23.00  & \bf 34.21\\
		\hspace{0.2cm} $recall$ & 23.04  & \bf 37.39\\
		\hspace{0.2cm} $fmeasure$ & 22.33  & \bf 34.57\\
		METEOR \cite{meteor} & 25.04   & \bf 41.75 \\\bottomrule
	\end{tabular} 
\end{table}

The statistical analyses presented in \tabref{tab:test-summarizer} validate STUNT's superior performance compared to IR-Jaccard. Notably, we observe significant $p$-values and medium effect sizes for BLEU-4 and ROUGE-LCS (f-measure), while METEOR demonstrates a large effect size.

\begin{table}[ht]
	\centering
	\caption{Statistical Tests: STUNT \emph{vs} IR-Jaccard}
	\scriptsize
	\label{tab:test-summarizer}
	\begin{tabular}{llrc}
		\toprule
		\textbf{Comparison} & \textbf{Metric} & \textbf{\emph{p}-value} & \textbf{d} \\ 
		\midrule
		\multirow{3}{*}{IR (Jaccard)  \emph{vs} STUNT} & BLEU-4 & $<$0.001 & -0.451 (M) \\ 
		& ROUGE-LCS (f-measure) & $<$0.001 & -0.471 (M) \\ 
		& METEOR & $<$0.001 & -0.474 (L) \\
%		\multirow{3}{*}{IR-2  \emph{vs} STUNT} & BLEU-4 & $<$0.001 & -0.71 (L) \\ 
%		& ROUGE-LCS (f-measure) & $<$0.001 & -0.72 (L) \\ 
%		& METEOR & $<$0.001 & -0.71 (L) \\
		\bottomrule
	\end{tabular}
\end{table}

While the metrics we computed provide a fair comparison among the experimented techniques, they do not give a clear idea of the quality of the summaries generated by STUNT. To this aim two of the authors manually inspected 384 randomly selected summaries generated by STUNT for which the generated text was different from the target summary (\ie the one written by developers). These are cases that in a ``binary quantitative evaluation'' would be classified as wrong predictions. The authors independently classified each summary as \emph{meaningful} or \emph{not meaningful}, based on the ability of the summary to properly describe the documented snippet. In the labeling, the two involved authors achieved a Cohen’s kappa \cite{cohen1960coefficient} of 0.61, indicating a substantial agreement when measuring inter-rater reliability for categorical items.

Conflicts, arisen in 71 cases and have been solved through open discussion among the authors. We classified 224 summaries as meaningful, with some of them representing even a better summary than the one manually written by the original developers. For example, we found the comment \texttt{if we have a frontend then we need to get the action list} to be more meaningful and detailed than the \texttt{exit if we do not have a frontend} written the developer. However, we also want to highlight the $\sim$41\% (160) of automatically generated summaries which were not meaningful and that stress how far we still are from obtaining a code summarizer being accurate enough to be deployed to developers (\ie generating correct summaries in most of cases).

% !TEX root = main.tex
%%%%%%%%%%%%%%%%%%%%%%%%%%%%%%%%%%%%%%%%
%%%%%%%%%%%%%%%%%%%%%%%%%%%%%%%%%%%%%%%%
\section{Threats to Validity} \label{sec:threats}
%%%%%%%%%%%%%%%%%%%%%%%%%%%%%%%%%%%%%%%%
%%%%%%%%%%%%%%%%%%%%%%%%%%%%%%%%%%%%%%%%

We discuss the threats that could affect the validity of our findings.\smallskip

\textbf{Internal Validity.} Building our dataset of classified and linked code comments (\secref{sec:dataset}) involved a certain degree of subjectivity. To partially address this threat, two evaluators independently assessed each instance and a third one solved conflicts when needed. Still, imprecisions are possible.

We performed a limited hyperparameters tuning of the T5 models, only experimenting with different learning rates. For example, we did not change the number of layers, but relied on the default T5$_{small}$ architecture by Raffel \etal \cite{raffel2019exploring}. Better results could be achieved with additional tuning. Also, relying on pre-trained code models like CodeT5 \cite{wang2021codet5}, might produce better results.

\textbf{Construct Validity.} When experimenting with SALOON, we compared its performance with the technique by Chen \etal \cite{chen2019automatically}. However, since their approach is not publicly available, we had to reimplement it following the paper's description.

We release our implementation \cite{replication}. Still related to the used baselines, as explained in \secref{sub:design_generation} we did not manage to compare STUNT (our approach for snippet summarization) with RL-BlockCom \cite{huang2020towards} on our dataset. At least, we presented a comparison performed on the dataset released by the authors.\smallskip

\textbf{External Validity.} The manually built dataset represents the obvious bottleneck in terms of generalizability, since it is based on the analysis of ``only'' 1,508 Java files and also capped our training/evaluation of SALOON. Still, building such a dataset costed over 815 man-hours. Also, we did not compare our technique against general purpose large language models such as ChatGPT \cite{chatgpt}, since designing a fair evaluation is challenging due to the unknown training set behind these LLMs. For example, we could have tested the ability of ChatGPT to summarize specific snippets which, however, were part of its training set together with their related comment.
% !TEX root = main.tex
%%%%%%%%%%%%%%%%%%%%%%%%%%%%%%%%%%%%%%%%
%%%%%%%%%%%%%%%%%%%%%%%%%%%%%%%%%%%%%%%%
\section{Related Work} \label{sec:related}
%%%%%%%%%%%%%%%%%%%%%%%%%%%%%%%%%%%%%%%%
%%%%%%%%%%%%%%%%%%%%%%%%%%%%%%%%%%%%%%%%

%Different papers in the literature highlighted the importance of commenting the code and the strong relationship between the comment and the commented code. Arafat and Riehle \cite{arafat2009commenting} showed that a best practice in high quality open source software is to strongly document the source code with comments, that also help in the maintenance of the project. Tan \etal \cite{tan2007icomment} performed a textual analysis of code comments that revealed a strong similarity between the code and the related comment, that can be enhanced to find inconsistencies.
We discuss techniques for (i) the automated linking of code to comments, and (ii) code summarization. 

\subsection{Linking Documentation to Code}
Some works, while studying code comments from different perspectives, came up with possible heuristics to identify the scope of code comments. Haouari \etal \cite{haouari2011good} showed that code comments frequently document the following code. 

While this is confirmed in our dataset, we also observed that it is far from trivial to assess the exact set of (following) lines actually documented by the comment due to the lack of a clear separator isolating the documented from the undocumented code. 

Fluri \etal \cite{Fluri:wcre07}, while studying the co-evolution of code and comments, suggested that token-based similarity between the code and the comment can be used to identify documented statements. Such an intuition has also been echoed by McBurney and McMillan \cite{mcburney2016empirical}. As shown in our study, our DL-based approach substantially outperforms similarity-based heuristics. 

Finally, Chen \etal \cite{chen2019automatically} recently proposed a machine-learning based method for the automatic identification of code comments scope. Such an approach has been extensively described in \secref{sub:designLinking} as one of the baselines we compared with. Our approach outperforms this approach as well.

%Chen \etal \cite{chen2019automatically} proposed a general method for the detection of source code comment scopes. They extracted three different features: (i) code features that describe a generic statement (\eg number of lines, and type of statement), (ii) comment features that define some characteristics of a comment (\eg number of lines of the comment), and (iii) relationship features that compute some similarity metrics between a snippet and a comment (\eg common keywords between a comment and a statement). They feed a random forest model with the three features in order to predict whether a defined comment is commenting each statement of the method. The model showed promising results, being able to correctly link 81\% of the comment to the referred statements. Therefore, we chose this model as a baseline for our experiments.

%\vspace{-0.5cm}
\subsection{Code Summarization}
Several techniques have been proposed to automatically summarize source code \cite{zhu2019automatic}. We focus our discussion on (i) techniques aimed at documenting code snippets (regardless of the underlying techniques used) and (ii) DL-based approaches (regardless of the target code granularity).

%\textbf{Variable} Sridhara \etal \cite{sridhara2011generating} proposed a heuristics that exploited syntactic and semantic information to automatically generate comments for the parameters of a specific method. These comments can be used by the developers as an overview that provide meaningful information about the parameters. A study with experienced developers showed that this overview is accurate.

Most of the techniques targeting the documentation of code snippets are based on IR. Representative works in this area are \emph{CodeInsight} \cite{Rahman:SCAM15}, \emph{ColCom} \cite{wong2015clocom, wong2013autocomment}, and ADANA \cite{aghaj:2019a}. %\emph{CodeInsight} \cite{Rahman:SCAM15} has been presented by Rahman \etal to extract from Stack Overflow discussions code comments relevant for a given code snippet. A similar idea has been exploited \emph{ColCom} \cite{wong2015clocom, wong2013autocomment} and in ADANA \cite{aghaj:2019a} which, however, look for clones of the snippet to document within a previously built code base of open source projects. 
The IR baselines we exploit is \secref{sub:design_generation} are representative of these works. 

Another family of techniques related to code snippets documentation relies on manually defined templates to describe high-level actions performed within functions. Seminal work in this area are from Sridhara \etal \cite{sridhara2011automatically} and Wang \etal \cite{wang2017automatically}. These approaches, while valuable, cannot generalize to all combinations of code statements one could expect to find since they are based on predefined templates. 
For this reason, data-driven techniques exploiting DL have been proposed \cite{zheng2017code,huang2020towards}. When it comes to snippet-level granularity, \emph{RL-BlockCom} \cite{huang2020towards} represents the state-of-the-art. As shown, our approach performs substantially better than \emph{RL-BlockCom}.

Most of the other DL-based techniques proposed in the literature focused on documenting entire functions. Liang \etal \cite{liang2018} presented Code-RNN, a Recursive Neural Network exploiting a GRU cell (Code-GRU) specifically designed for code comments generation. The authors show that their approach can achieve higher ROUGE score \cite{lin2004rouge} as compared to vanilla DL models not tailored for source code. Hu \etal \cite{Hu:icpc2018} built a dataset of $\langle$method, javadoc$\rangle$ pairs from $\sim$9k Java projects to train a Deep Neural Network (DNN) aimed at documenting Java methods. The authors used the BLEU-4 score \cite{Papineni:2002} to compare the summaries generated by their approach to those of the neural attention model by Iyer \etal \cite{iyer:acl}, showing the superiority of the proposed technique. 

While previous works represented code as a stream of tokens, other authors combined such a representation with one capturing AST information \cite{Wan:ase2018,LeClair:icse2019}. For example, LeClair \etal \cite{LeClair:icse2019} showed how exploiting AST-based information allows to improve the performance achieved by both Hu \etal \cite{Hu:icpc2018} and Iyer \etal \cite{iyer:acl}. The work by LeClair \etal has been later on extended and improved by Haque \etal \cite{haque:2020}, which provide as input to the model additional information related to the ``file context'' of the method to summarize. They show that such a contextual information helps to further boost performance.
%Two recent works by Mastropaolo \etal \cite{mastropaolo2021studying} and Lin \etal \cite{lin2021improving} reported new state-of-the-art performance for method-level code summarization by using Transformer models. In our paper, we exploit the same T5 architecture used in \cite{mastropaolo2021studying} by applying, however, a different training procedure aimed at documenting code snippets rather than entire methods.

Zhang \cite{Zhang:icse2020} showed that combining IR and DL techniques it is possible to boost the performance of function-level code summarization. Our work focuses on the related but different problem of snippet summarization that, as explained, poses different challenges especially in the building of the training data.
% !TEX root = main.tex
%%%%%%%%%%%%%%%%%%%%%%%%%%%%%%%%%%%%%%%%
%%%%%%%%%%%%%%%%%%%%%%%%%%%%%%%%%%%%%%%%
\section{Conclusions} \label{sec:conclusions}
%%%%%%%%%%%%%%%%%%%%%%%%%%%%%%%%%%%%%%%%
%%%%%%%%%%%%%%%%%%%%%%%%%%%%%%%%%%%%%%%%

We targeted the problem of code snippet summarization, presenting (i) a manually labeled dataset of $\sim$6.6k code comments classified in terms of information they provide (\eg code summary) and linked to the code statements they document; (ii) SALOON, a T5 model trained on our manually built dataset to automatically classify and link inner comments in Java code; and (iii) STUNT, a T5 model trained on a large-scale dataset of documented code snippets automatically created by running SALOON on 10k Java projects. 

We achieved promising results for both code linking and snippet summarization, pointing however to the need for  research in this field. Our dataset and our models, publicly released \cite{replication}, represent a step in that direction.

\section*{Acknowledgment}
This project has received funding from the European Research Council (ERC) under the European Union's Horizon 2020 research and innovation programme (grant agreement No. 851720). 

\bibliographystyle{ACM-Reference-Format}
\bibliography{main}

\end{document}